\documentstyle{article}
\begin{document}
\pagestyle{empty}
\begin{flushright}
Groningen, 23rd February, 1995.
\end{flushright}

\begin{center}
\large Gravitational redshift in the NGC~4258  disk?
\end{center}

SIR. The very interesting observations by Myoshi et al. (1995) of
water-vapour maser emission in NGC~4258 offer an additional direct
test for their model of a high rotation velocity disk and allow an
independent determination of the central black hole mass. Indeed, a
detectable amount of gravitational redshift should affect the observed
line wavelengths.

As the distances of the maser emission regions from the center of the
black hole $r$ are much larger than the black hole Schwarzschild
radius $r_{\rm S} = 2 G M/c^2 = 3.5 ~10^{-6}~{\rm pc}$, the
gravitational redshift is simply given by (assuming a spherical symmetry):
\begin{equation}
 z(r) = \frac{r_{\rm S}}{2 ~ r} = \frac{v_{\rm circ}^2}{c^2},
  \nonumber
  \label{redshift}
\end{equation}
where $G$ is the gravitational constant, $M$ is the mass of the
central object, $v_{\rm circ}$ is the velocity of a particle in
Keplerian motion around it and $c$ is the velocity of light. The
gravitational redshift expected for a central mass of $3.6~10^{7} ~
{\rm M}_{\odot}$ at the inner edge of the disk is $c z(r=0.13 {\rm pc})
= 4.0 ~ {\rm km}~{\rm s}^{-1}$, while the one at the outer edge is $c
z(r=0.26 {\rm pc}) = 2.0 ~ {\rm km}~{\rm s}^{-1}$, i.e. more than ten
times their velocity resolution, or about 4 and 2 times their quoted
errors on the velocity determination for the high-velocity lines.

The high-rotation velocity lines offer a easy and safe way to test the
disk model and to determine the black hole mass.  The reason is that
they apparently arise very close to a line perpendicular to the
line-of-sight. As a consequence, their velocities should
systematically deviate from the values expected from Keplerian
motion, with residuals that are inversely proportional to the
distances of the maser emission regions from the center of the black
hole. Furthermore, the Keplerian motion velocities have an
anti-symmetric behaviour relative to the center of the disk, while the
gravitational redshifts have a symmetric one. The
low-velocity lines should also be affected by a gravitational redshift
of $4.0 ~ {\rm km}~{\rm s}^{-1}$ as they are
produced at the inner edge of the disk ($r = 0.13 {\rm pc}$). \\

\noindent Alain Smette\\
Konrad Kuijken\\
Kapteyn Astronomical Institute\\
P.O. Box 800\\
NL-9700 AV Groningen\\
The Netherlands\\

{\bf Reference:}
Myoshi, M., Moran, J., Herrnstein, J., Greenhill, L., Nakai, N.,
Diamond, P. \& Inoue, M. {\it Nature \/} {\bf 373}, 127 (1995)

\end{document}